\documentclass[a4paper,fleqn,usenatbib]{mnras}

\usepackage[T1]{fontenc}
\usepackage{ae,aecompl}

\usepackage{graphicx}	
\usepackage{amsmath}	
\usepackage{amssymb}	
\usepackage{multicol}        
\usepackage{bm}		
\usepackage{pdflscape}	

\title[Stellar Mixing Impact on the Vertical Metallicity Gradient]{Impacts of a Flaring Star-forming Disc and Stellar Radial Mixing on the Vertical Metallicity Gradient}

\author[D. Kawata et al.]{
Daisuke Kawata,$^{1}$\thanks{E-mail: d.kawata@ucl.ca.uk}
Robert J. J. Grand,$^{2,3}$
Brad K. Gibson,$^{4}$
Luca Casagrande,$^{5}$
\newauthor{Jason A. S. Hunt$^{1}$ and Chris B. Brook$^{6,7}$}
\\
$^{1}$Mullard Space Science Laboratory, University College London, Holmbury St. Mary, Dorking, Surrey, RH5 6NT, UK\\
$^{2}$Heidelberger Institut f\"{u}r Theoretische Studien, Schloss-Wolfsbrunnenweg 35, 69118 Heidelberg, Germany\\
$^{3}$Zentrum f\"{u}r Astronomie der Universit\"{a}t Heidelberg, Astronomisches Recheninstitut, M\"{o}nchhofstr. 12-14, 69120 Heidelberg, Germany\\
$^{4}$E.A. Milne Centre for Astrophysics, Dept of Physics \& Mathematics, University of Hull, Hull, HU6 7RX, United Kingdom \\
$^{5}$Research School of Astronomy \& Astrophysics, Mount Stromlo Observatory, The Australian National University, ACT 2611, Australia\\
$^{6}$Departamento de F\'isica Te\'orica, Universidad Aut\'onoma de Madrid, 28049 Cantoblanco, Madrid, Spain\\
$^{7}$Astro-UAM, UAM, Unidad Asociada CSIC\\
}

\date{Accepted XXX. Received YYY; in original form ZZZ}

\pubyear{2016}

\begin{document}
\label{firstpage}
\pagerange{\pageref{firstpage}--\pageref{lastpage}}
\maketitle

\begin{abstract}
Using idealised N-body simulations of a Milky Way-sized disc galaxy, we qualitatively study how the metallicity distributions of the thin disc star particles are modified by the formation of the bar and spiral arm structures. The thin disc in our numerical experiments initially has a tight negative radial metallicity gradient and a constant vertical scale-height. We show that the radial mixing of stars drives a positive vertical metallicity gradient in the thin disc. On the other hand, if the initial thin disc is flared, with vertical scale-height increasing with galactocentric radius, the metal poor stars originally in the outer disc become dominant in regions above the disc plane at every radii. This process can drive a negative vertical metallicity gradient, which is consistent with the current observed trend. This model mimics a scenario where the star-forming thin disc was flared in the outer region at earlier epochs. Our numerical experiment with an initial flared disc predicts that the negative vertical metallicity gradient of the mono-age relatively young thin disc population should be steeper in the inner disc, and the radial metallicity gradient of the mono-age population should be shallower at greater heights above the disc plane. We also predict that the metallicity distribution function of mono-age young thin disc populations above the disc plane would be more positively skewed in the inner disc compared to the outer disc.
\end{abstract}




\section{Introduction}
\label{sec:intro}

Both the Milky Way \citep{yy82,gilr83} and external galaxies \citep[e.g.][]{db79,db02} show geometrically thick discs in addition to the geometrically thin disc. Traditionally, the thick and thin disc components are separated by fitting stellar density distributions (or surface brightness for external galaxies) with two different scale-heights. For example, using the photometric data of the Sloan Digital Sky Survey (SDSS), \citet{jibls08} analysed the stellar number density map for the Milky Way disc. They demonstrated that the observed stellar number density distribution were well fitted by the two (thin and thick) disc components with scale-heights of 300 pc and 900 pc respectively. We call this definition of thick and thin discs "geometrically" thick and thin discs. 

High-resolution spectra of kinematically selected nearby geometrically thick disc stars show systematically higher $\alpha$-element abundance with respect to the iron abundance, [$\alpha$/Fe], compared to kinematically colder geometrically thin disc stars with similar [Fe/H] \citep[e.g.][]{pncmw00,fbl03}. There is a trend that the (kinematically selected) geometrically thick disc stars follow the high-$\alpha$ sequence, while the geometrically thin disc stars follow the distinct low-$\alpha$ sequence. However, the recent studies of a larger number of nearby dwarf stars showed that kinematically selected geometrically thick (thin) disc stars do not always follow the high-$\alpha$ (low-$\alpha$) sequence \citep[e.g.][]{assdg12}, and the phase-space distribution of the high-$\alpha$ and low-$\alpha$ disc stars overlaps significantly with each other \citep[e.g.][]{bfo14}. Therefore, these studies suggest that we should define thick and thin disc stars depending on their chemical properties \citep[e.g.][]{lbaij11,navfa11}, although it requires high-quality high-resolution spectroscopic survey which is observationally quite expensive. The chemically defined thick and thin discs are not necessarily the same as geometrically defined thick and thin discs. Interestingly, the chemically defined thick disc is found to be radially smaller than the chemically defined thin disc \citep[e.g.][]{baboym11,brlhbl12,crmlb12}, which may indicate a difference in their formation epoch \citep{bkmg06}.

 Comparing the spectroscopically derived stellar parameters with theoretical isochrones, \citet{hdmlkg13} analysed the stellar ages, and separated their sample of stars by chemical abundance (e.g.\ high-$\alpha$ and low-$\alpha$ discs) and by age. In general, the high-$\alpha$ disc stars are older than low-$\alpha$ disc stars \citep[see also][]{brsfa14,mg15}, although some "anomalies"  of young high-$\alpha$ disc stars are also found \citep{carmm15,mrahm15}. Therefore, in this paper, we consider that high-$\alpha$ disc stars are generally older than the low-$\alpha$ disc stars, and call the young (old) disc population the "thin (thick) disc population" or "thin (thick) disc component".
  
 By dividing the sample of the disc stars into thin and thick disc populations by low-$\alpha$ and high-$\alpha$ stars from the Gaia-ESO survey, \citet{mhrbdv14} analysed the radial metallicity gradients of the thin and thick disc populations within the region of about $6<R<12$ kpc and $|z|<0.6$ kpc. They also measured the vertical metallicity gradients of the thin and thick populations within the region of about $|z|<2$ kpc and $7<R<9$ kpc. For the thin disc population of their "main sample" of 341 stars with a S/N higher than 15 per pixel,  \citet{mhrbdv14} obtained $d\text{[M/H]}/dR=-0.045\pm0.012$ and $d\text{[M/H]}/dz=-0.079\pm0.013$, where M indicates the total metal abundance. In other words, the thin disc population has a negative radial metallicity gradient and negative vertical metallicity gradient. On the other hand, they found that the thick disc population shows no radial metallicity gradient, but a very shallow negative vertical metallicity gradient. This indicates that the chemical properties of the thick disc population is radially well mixed  \citep[see also][]{baboym11,nbbah14}, while the thin disc population preserves their radial and vertical metallicity gradient well. 
 
 The clear existence of a radial metallicity gradient is an important constraint on the evolutionary history of the Galactic thin disc, because radial mixing of stars is found in the previous studies \citep[e.g.][]{kpa13,mcm14} to significantly flatten the radial metallicity gradient of the old thin disc population\footnote{This can apply only if there is a significant radial metallicity gradient. It is still in debate if there were a negative radial metallicity gradient in star forming disk especially at a high redshift when strong feedback could flatten the radial metallicity gradient \citep{pfg12,gpbsb13}.}
However, some of these model predictions are based on the results of simulations with a strong bar which appears larger than what is seen in the Milky Way. Using a simulation with a more reasonably sized bar for the Milky Way, and including the gas, star formation and chemical evolution, \citet{gkc15} showed that the stellar radial metallicity gradient does not change significantly. On the other hand, the dispersion of the metallicity distribution function (MDF) increases, which is qualitatively consistent with the observed MDF for different ages of stars in the solar neighbourhood \citep{csacrmf11}. \citet{gkc15} also showed that the radial metallicity gradient of the gas component remains steep, because of the higher star formation activities in the inner region of the disc and the efficient local metal diffusion for the gas which also suffers from radial mixing.

Compared to the radial metallicity gradient, the impact of the radial mixing on the vertical metallicity gradient for the thin disc population is not well studied \citep[for studies for the combined populations of thin and thick discs, see e.g.][]{cknn11,lrdiqw11,pfg12,rdl13,mpgbsb16,rspm16}. \citet{mcm13} discussed the vertical metallicity gradients for the populations with different ages in their hybrid chemical evolution model adding dynamical information with post-processing cosmological disc galaxy formation simulation data. They presented that the vertical metallicity gradients of the sample of disc populations with a similar age, i.e. "mono-age" disc population, are almost flat. This seems consistent with the  negligible vertical metallicity gradients for different samples of stars with similar [$\alpha$/Fe] abundance ratios from SDSS/Sloan Extension for Galactic Understanding and Exploration (SEGUE) data \citep{sjrlb14}, if [$\alpha$/Fe] is considered to be related to the age. However, the simulations of \citet{mcm13} seem to suffer from too much radial mixing of stars, because they predicted that the "metallicity distributions of (unbiased) samples at different distances from the Galctic centre peak at approximately the same value, [Fe/H]$\sim-0.15$ dex" \citep{mcm14}, which is inconsistent with the recent results of the SDSS/Apache Point Observatory Galactic Evolution Experiment (APOGEE) survey \citep{hbhnb15}\footnote{It is worth mentioning that using a self-consistent cosmological simulation \citet{ldnhh16} has successfully reproduced the observed trend of the different shapes of the MDF in the different radial and vertical regions of the Galactic disc in \citet{hbhnb15}.}. Therefore, we think that it would be useful to have more simplified numerical experiments to study the impact of the radial mixing on the vertical metallicity gradients of the mono-age thin disc populations.

As mentioned above, the thin disc population as a whole shows a clear negative vertical metallicity gradient \citep{mhrbdv14}. \citet{vmtb06} claimed an even steeper vertical metallicity gradient of $d\text{[M/H]}/dz=-0.58\pm0.13$ for the stars younger than 4 Gyr. Using data with more accurate age measured with Kepler asteroseismic data in \citet{csassh16}, we have measured the vertical metallicity gradient of the stars within the age range of $0-3$ Gyr and $3-5$ Gyr, and obtained a negative gradient of $-0.29$ and $-0.09$ dex~kpc$^{-1}$ respectively. Please note that the number of stars in each sample is small, and so the quoted gradients should be approached with some caution. Still, this result infers that each mono-age thin disc population has a negative vertical metallicity gradient, although our age range may be still too large to call it a mono-age population. On the other hand, the thin disc population stars are expected to form within very thin high density molecular gas. For example, the recent measurement of the metallicity of a nearby sample of B stars show a very small dispersion \citep[e.g.][]{mfnnp12,sfmc13}. It is a natural expectation that the star forming gas disc has almost no vertical metallicity gradient. Then, how was the observed negative vertical metallicity gradient of mono-age thin disc populations built up?

Using simple numerical experiments based on N-body simulations of an idealised isolated galactic disc, we qualitatively study the impact of the radial mixing on the vertical metallicity gradient of the thin disc population. In this paper, we analyse the radial and vertical metallicity gradient using the current position of the particles. We use "radial mixing" to describe the overall radial re-distribution due to both "churning" and "blurring" \citep{sb09a,rspm16}. Churning indicates the change of angular momentum of stars, and blurring describes the radial re-distribution of stars due to their epicyclic motion. Churning can split into the two mechanisms which we call "radial scattering" and "co-rotation radial migration". Radial scattering describes change of angular momentum with kinematic heating, i.e. the increase in random energy of the orbit. On the other hand, co-rotation radial migration indicates gain or loss of angular momentum of stars at the co-rotation resonance of the bar or spiral arms, which does not involve any kinematic heating \citep{jsjb02}. In this paper, we discuss the effect of radial mixing, without distinguishing these mechanisms, unless otherwise specified. Section~\ref{sec:meth} describes briefly the numerical simulation code and numerical models. In Section~\ref{sec:resfm}, we first demonstrate that a mono-age population disc initially with a constant vertical scale-height and a tight negative radial metallicity gradient develops a positive vertical metallicity gradient after the radial mixing of the stars due to interaction with the bar and spiral arms. In Section~\ref{sec:resfldm}, we present one of the possible solutions to develop and keep a negative vertical metallicity gradient for a mono-age thin disc population, which is a scenario where the mono-age population formed in a flaring star-forming disc. A summary and discussion of this study is presented in Section~\ref{sec:sum}.

\section{Method and Models}
\label{sec:meth}

We use our original Tree N-body code, {\tt GCD+} \citep{kg03a,kogbc13} for the numerical experiments presented in this paper, and model the evolution of a barred disc galaxy similar in size to the Milky Way. We initially set up an isolated disc galaxy which consists of stellar discs, with no bulge component, in a static dark matter halo potential \citep{rk12,gkc12a}. A live dark matter halo can respond to the disc particles by exchanging angular momentum. However, if not properly modelled, a live dark matter halo may introduce some numerical scattering and heating \citep[e.g.][]{dvh13}. In the interest of a more controlled experiment and computational speed, we use a static dark matter halo.
 We use the standard Navarro-Frenk-White (NFW) dark matter halo density profile \citep{nfw97}, assuming  a $\Lambda$-dominated cold dark matter ($\Lambda$CDM) cosmological model with cosmological parameters of $\Omega_0=0.266=1-\Omega_{\Lambda}$, $\Omega_{\rm b}=0.044$, and $H_0=71{\rm kms^{-1}Mpc^{-1}}$:

\begin{equation}
\rho _\text{dm}=\frac{3H_{0}^{2}}{8\pi G}\frac{\Omega _{0}-\Omega_b}{\Omega_0}\frac{\delta _{c}}{cx(1+cx)^{2}},
\end{equation}
where
\begin{equation}
c=\frac{r_{200}}{r_\text{s}}, \;\; x=\frac{r}{r_{200}},
\end{equation}
and
\begin{equation}
r_{200}=1.63\times 10^{-2}\left(\frac{M_{200}}{h^{-1}M_{\odot }}\right)^{\frac{1}{3}} h^{-1}\textup{kpc},
\end{equation}
\noindent where $\delta _{c}$ is the characteristic density of the profile \citep{nfw97}, $r$ is the distance from the centre of the halo and $r_{s}$ is the scale radius. The total halo mass is set to be $M_{200}=1.5\times 10^{12}M_{\odot }$ and the concentration parameter is set at $c=12$. The halo mass is roughly consistent with  the recent estimated mass of the Milky Way \citep[e.g.][]{pjm11}. 

The stellar disc is assumed to follow an exponential surface radial density profile and $\textup{sech}^{2}$ vertical density profile:
\begin{equation}
\rho _\text{d}=\frac{M_\text{d}}{4\pi z_\text{d}R_\text{d}^2}\textup{sech}^{2}\left(\frac{z}{z_\text{d}}\right)\exp \left(-{\frac{R}{R_\text{d}}}\right),
\end{equation}
\noindent where $M _\text{d}$ is the stellar disc mass, $R_\text{d}$ is the scale length and $z_\text{d}$ is the scale height. In this paper, we focus on the results of the simulations of two different models. A fiducial model, Model A, consists of two stellar disc components, which we call thick and thin discs. These two discs are intended to mimic a radially smaller chemically defined thick population and a radially larger chemically defined thin disc population discussed in Section~\ref{sec:intro}. In the scenario discussed in Section~\ref{sec:intro}, we expect the thin disc population grows over 8 Gyr, after the thick disc population formed. However, in this simple numerical experiment, we ignore the mass growth of the disc and only focus on the structure changes driven by the development of non-axisymmetric structures, such as a bar and spiral arms. The assumed mass, scale length and scale heights for each disc component are shown in Table~\ref{tab:model}. The other model, Model B, has a 2nd thin disc component, thin2 in Table~\ref{tab:model}, which is thinner than the thin disc of Model~A in the inner region, and the scale height changes as a function of radius, as follows
\begin{equation}
z _\text{d}(R)=z_{d,0} \exp{(R/R_\text{fl})},
\end{equation}
where $z_{d,0}$ is the scale height at $R=0$, and $R_\text{fl}$ is the scale length, which controls the degree of flaring. The assumed parameters for Model B are also shown in Table~\ref{tab:model}.

For Model A, we use 4,500,000 and 1,500,000 particles for the thin and thick disc components of the initial condition respectively. This leads to 10,000 M$_{\odot}$ for each particle. Our choice of a relatively high mass of the thick disc component was inspired by \citet{shdmlckg14}. Model~B also uses the same particle mass resolution, and 500,000 particles for the 2nd thin disc component. We apply the spline softening suggested by \citet{pm07}. We set a softening length of 341 pc (the equivalent Plummer softening length is about 113 pc). To relax the initial system without kinematic heating from non-axisymmetric structures, we ran a simulation for 2 Gyr (4 Gyr) for Model A (Model B), randomly moving randomly selected particles azimuthally, which suppresses the development of non-axisymmetric structures \citep{pjmwd07}. We then run the simulations for 8 Gyr which roughly corresponds to the age of the thin disc of the Milky Way \citep[e.g.][]{hdmlkg13}. Note that the time of $t=0$ Gyr in this paper indicates the starting time of this 8 Gyr simulation using the the end-product of the relaxing run as an initial condition. 

To make sure that there is no numerical heating affecting the structure evolution, we ran a test simulation with the same initial condition as Model B for 8 Gyr while randomly moving randomly selected particles azimuthally. In this case, no bar or spiral arms developed, and we have confirmed that the radial or vertical structure (even for the very thin inner region of the thin2 disc component) was hardly changed after 8 Gyr. Therefore, the resolution of our numerical models is high enough to resolve the vertical structure of the thin discs, and the results in this paper are very unlikely to be affected by the numerical heating. 

\begin{figure*}
	\includegraphics[width=0.8\hsize]{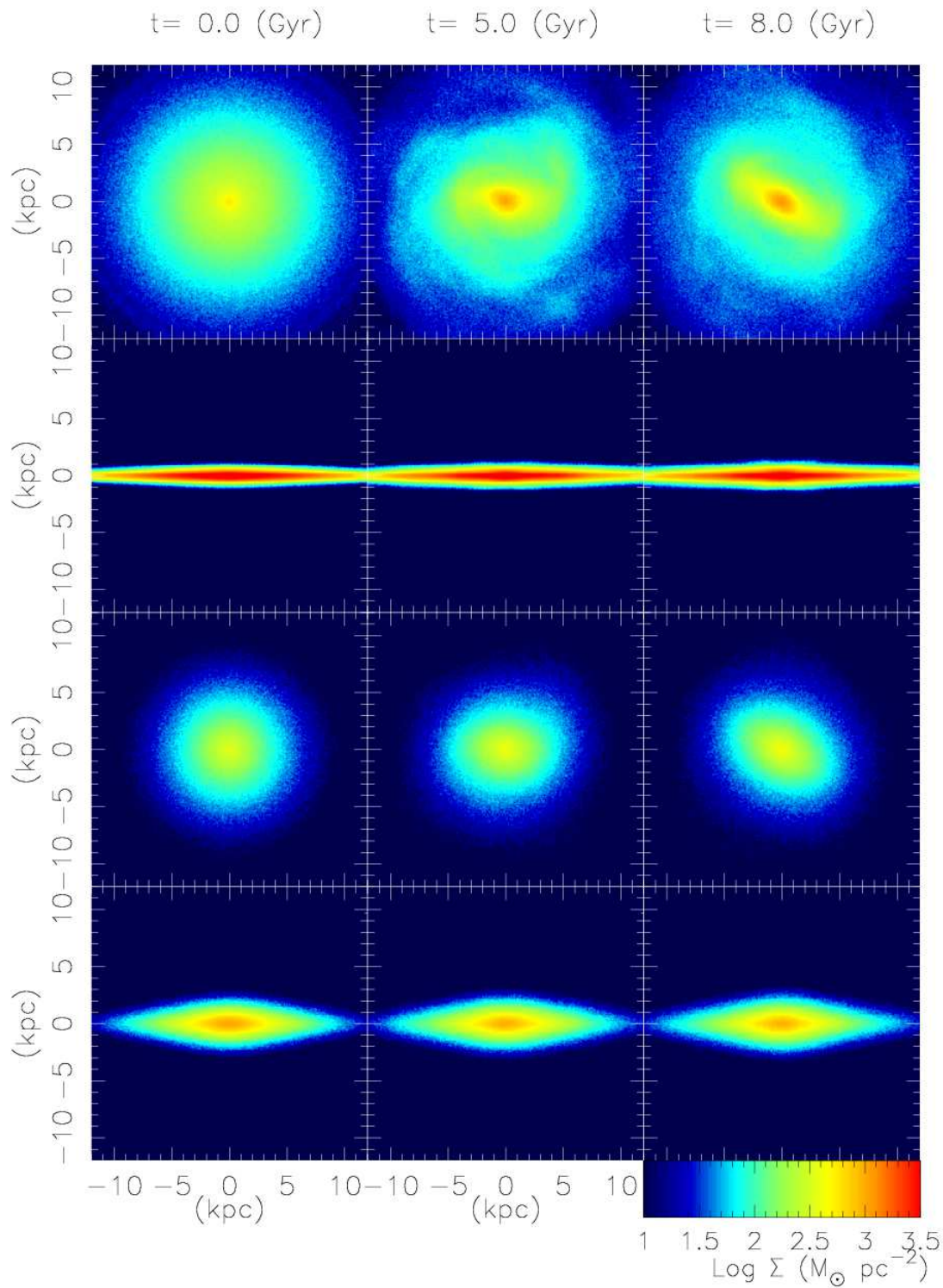}
    \caption{Snapshots of Model A which show the face-on (top panels) and edge-on (2nd panels) images of the thin disc component and the face-on (3rd panels) and edge-on (bottom panels) images of the thick disc components at $t=0$ (left), 5 (middle) and 8 (right) Gyr. }
    \label{fig:snap}
\end{figure*}

\begin{table*}
 \caption{Model parameters.}
 \label{tab:model}
 \begin{tabular}{lcccccccccccc}
  \hline
  Model & \multicolumn{3}{c}{Thick} & & \multicolumn{3}{c}{Thin} & & \multicolumn{4}{c}{2nd Thin} \\
  \cline{2-4} \cline{6-8} \cline{10-13}
    & $M_\text{d,thick}$ & $R_\text{d,thick}$ & $z_\text{d,thick}$ & & $M_\text{d,thin}$ & $R_\text{d,thin}$ & $z_\text{d,thin}$ & & $M_\text{d,thin2}$ & $R_\text{d,thin2}$
    & $z_\text{d,0,thin2}$ & $R_\text{fl}$ \\
   & $M_{\sun}$ & kpc & kpc & & $M_{\sun}$ & kpc & kpc & & $M_{\sun}$ & kpc & kpc & kpc \\
  \hline
  A & $1.5\times10^{10} $ & 2.5 & 1.0 & & $4.5\times10^{10}$ & 4.0 & 0.35 & & $-$ & $-$ & $-$ & $-$ \\
  B & $1.5\times10^{10} $ & 2.5 & 1.0 & & $4.0\times10^{10}$ & 4.0 & 0.35 & & $5\times10^{9}$ & 4.0 & 0.06 & 5.0 \\
  \hline
 \end{tabular}
\end{table*}

Fig.~\ref{fig:snap} shows the evolution of Model A. Model A fully develops the central bar by $t=5.0$ Gyr. Fig.~\ref{fig:snap} shows that the thick disc component also develops a bar, although it is rounder than the thin disc one. This is consistent with what is shown in \citet{bt11}.

\begin{figure*}
	\includegraphics[width=\hsize]{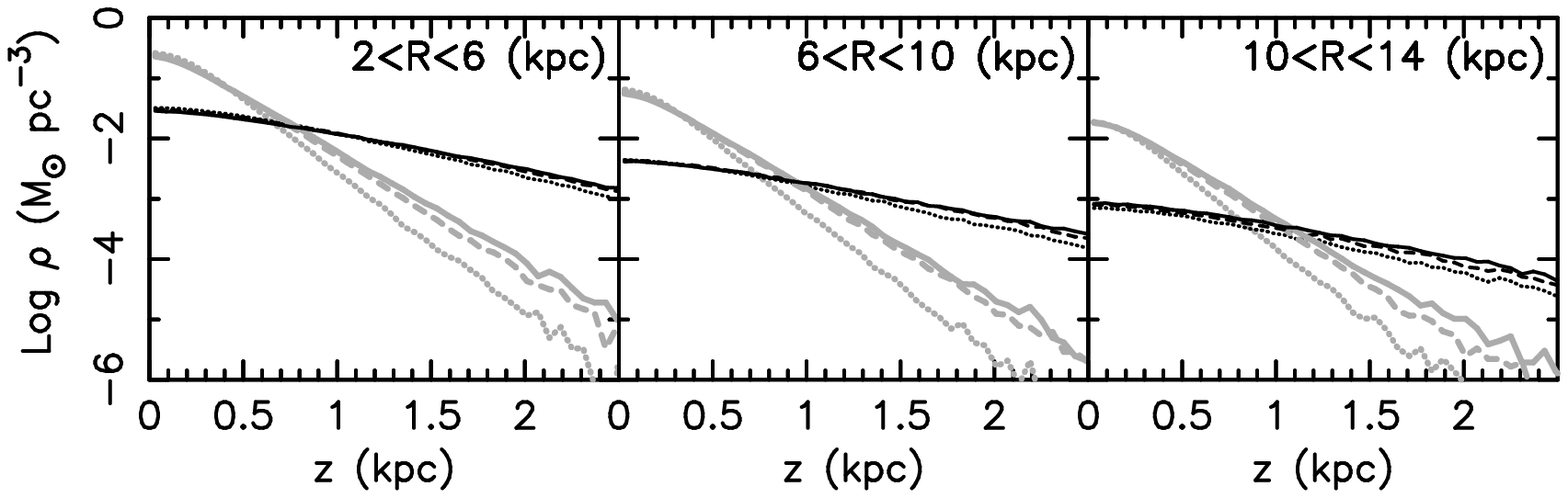}
    \caption{Vertical density profiles for the thick (black lines) and thin (thick grey lines) disc components of Model A at $t=0$ (dotted line), 5 (dashed line) and 8 (solid lines) Gyr. The left, middle and right panels show the density profiles in the radial ranges of $2<R<6$~kpc, $6<R<10$~kpc and $10<R<14$~kpc respectively. }
    \label{fig:zprof-m68}
\end{figure*}

Fig.~\ref{fig:zprof-m68} shows the evolution of the vertical density profiles of the thick and thin disc components of Model A in different radial ranges.  In this paper, we focus on the three different radial ranges, $2<R<6$ kpc, $6<R<10$ kpc and $10<R<14$ kpc. The vertical density profile of the thick disc component does not change significantly, despite the development of the bar. On the other hand, the thin disc component is heated up vertically, and the disc becomes thicker after the bar formation at $t=5.0$ Gyr. After the bar is fully formed, there is less significant heating happening. 

In this paper, we are interested in how the vertical metallicity gradients of a mono-age thin disc population are affected by the radial mixing of the stars induced by a bar and spiral arms. We focus on the thin disc component for Model~A and the 2nd thin disc component of Model~B. Unless otherwise specified, we discuss the radial and vertical metallicity distribution at $t=8$~Gyr, when we have tagged the metallicity at $t=0$~Gyr for the thin disc particles for Model~A and the 2nd thin disc particles for Model~B to follow the radial metallicity gradient of $\text{[M/H]}=0.5-0.08\times~R$ with a dispersion of 0.05~dex.  In Section \ref{sec:resfm}, we also present a result of Model~A5 where we use Model~A, but set the same metallicity gradient and dispersion at $t=5$~Gyr to see the evolution from $t=5$ to 8 Gyr after the bar fully developed. Our assumed radial metallicity gradient is similar to what is shown in Fig. 5 of \citet{sb09a} who compiled the observed metallicities of HII regions in the Milky Way and obtained the best-fit slope of $-0.082$~dex~kpc$^{-1}$. Our assumed metallicity dispersion of 0.05~dex was inspired by a very small metallicity variations observed for a nearby sample of B stars \citep[e.g.][]{mfnnp12,sfmc13}. These numerical experiments mimic the evolution of a mono-age thin disc population formed at 8 Gyr ago, and allow us to explore how the metallicity distributions of the mono-age population are changed by the radial mixing caused by the bar and spiral arms after 8~Gyr of evolution. Note that as mentioned above, we ignore the mass growth of the disc component to simplify the numerical experiments, which helps us to understand the results more clearly.

To simplify the discussion, we present here the results of only one initial metallicity gradient, i.e. $-0.08$ dex~kpc$^{-1}$, with zero vertical metallicity gradients for only two models, the thin disc component of Model~A and the thin2 disc component of Model~B. We have explored different N-body numerical experiments, e.g. with a lighter or no thick disc, more compact thick disc, or models with no bar, and found qualitatively consistent results with what are discussed in this paper for our Model~A. There is also a freedom in the choice of the initial radial metallicity gradients at $t=0$~Gyr. However, by studying cases with different initial metallicity gradients, we confirmed that the qualitative discussion in this paper is not affected by the values of the assumed initial metallicity gradient, as long as the initial radial metallicity gradient is negative. We therefore used an arbitrary negative radial metallicity gradient. The zero vertical metallicity gradient is chosen for simplicity, but it is a reasonable assumption, because for example in the Milky Way the metallicity dispersion of young B stars is observed to be very small \citep[e.g.][]{mfnnp12,sfmc13} and the young stars and star-forming regions are confined in the disc plane, which leads to a chemically well mixed interstellar medium in the vertical direction, and no vertical metallicity gradient in the star-forming disc.

In this paper, we measure the vertical metallicity gradients of the thin disc stars within the region of about $|z|<2$ kpc and $6<R<10$ kpc, following \citet{mhrbdv14}. We choose a wider radial range than the radial range observed in \citet{mhrbdv14} to increase the sample of the particles. However, for the qualitative discussion in this paper, this should not be an issue. We also analyse the vertical metallicity gradient in the radial range of $2<R<6$ kpc and $10<R<14$ kpc, where soon the ongoing \textit{Gaia} and ground-based spectroscopic surveys will provide data for the Milky Way. 

\section{Results}
\label{sec:res}

In Section~\ref{sec:resfm}, we first show the results of our fiducial model, Model~A, and demonstrate that it is difficult to keep the negative vertical metallicity gradient which is observed in the Milky Way, if there is radial mixing of the stars as commonly seen in N-body simulations. Then, in Section~\ref{sec:resfldm}, we will show one of the possible solutions to develop and keep the negative vertical metallicity gradient. 

\subsection{Fiducial Model}
\label{sec:resfm}

\begin{figure}
	\includegraphics[width=\hsize]{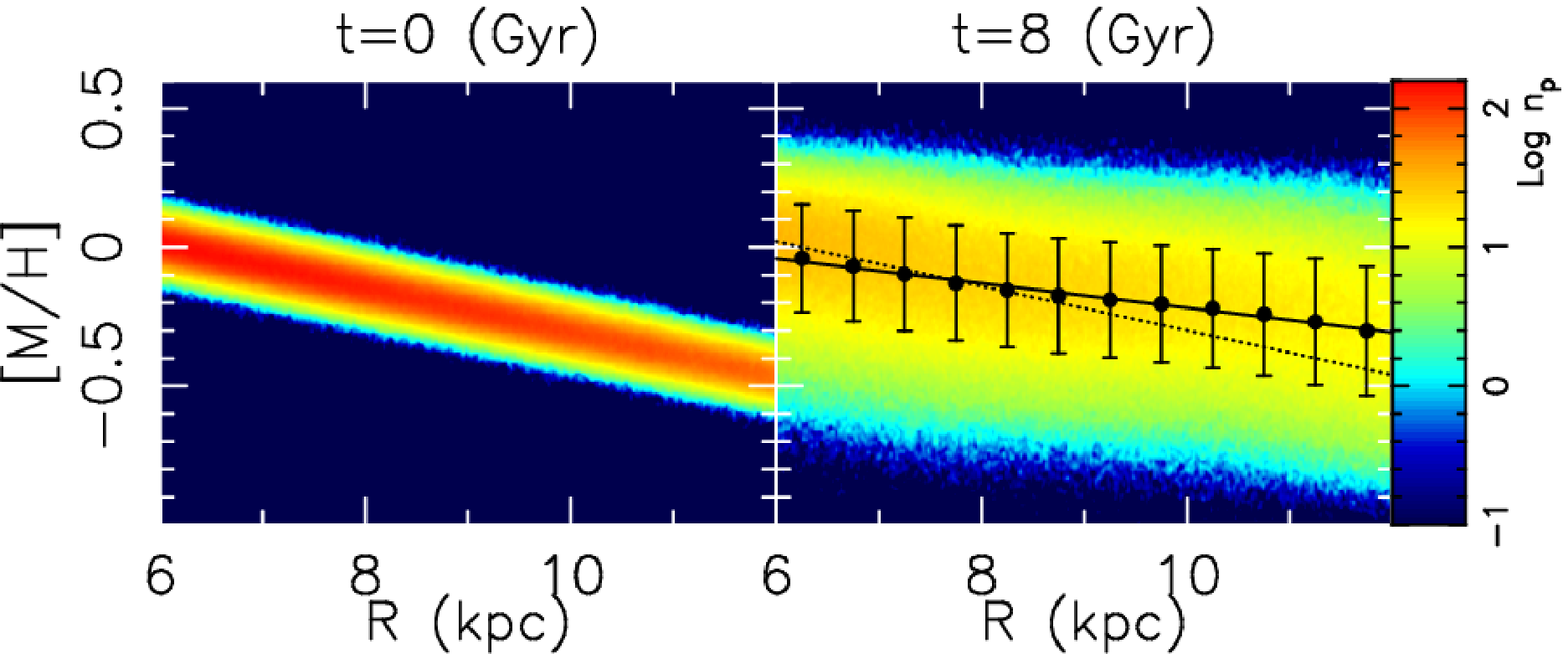}
    \caption{Radial metallicity gradient of the thin disc component for Model~A at $t=0$ (left) and 8 (right) Gyr. The black dots and the error bars in the right panel shows the mean and dispersion of [M/H] at different radii, and the solid line show the best fit straight line. The dotted line shows the mean metallicity gradient assumed at $t=0$ Gyr. Colour contour map indicates the number density of particles in arbitrary logarithmic units.}
    \label{fig:smaprmet-m68}
\end{figure}

Fig.~\ref{fig:smaprmet-m68} shows the radial metallicity distribution for the thin disc particles of Model~A at $t=0$ and 8~Gyr. To quantify the degree of radial mixing, following \citet{ldnhh16}, we measured the fraction of particles whose $|R(t=8)-R(t=0)|>2$~kpc among the particles within $6<R<10$ kpc at $t=8$~Gyr, and obtained 44~\%. This fraction indicates a significant radial mixing, but less significant than 55~\% seen in a cosmological simulation of \citet{ldnhh16}.

The evolution of the metallicity gradient is summarised in Table~\ref{tab:zgrad}. We analyse the radial metallicity gradient of the thin disc stars within the region of about $6<R<12$~kpc and $|z|<0.6$~kpc, following \citet{mhrbdv14}. The radial metallicity gradient becomes significantly shallower. However, the metallicity gradient does not become completely flat. We found that the radial metallicity gradient became shallower quickly till $t=2$~Gyr when $d\text{[M/H]}/dR=-0.057\pm0.028$, and then changed very slowly. The dispersion of the metallicity distribution at each radius becomes significantly higher, because of the radial scattering due to the bar formation and the co-rotation radial migration driven by the bar and spiral arms. This is consistent with a study of a shorter time evolution of  \citet{gkc15}. 

\begin{table*}
 \caption{Radial and vertical metallicity gradients.}
 \label{tab:zgrad}
 \begin{tabular}{lcccccccccc}
  \hline
  Model & \multicolumn{2}{c}{Time} & & \multicolumn{2}{c}{Initial} & & \multicolumn{4}{c}{final} \\
  \cline{2-3} \cline{5-6} \cline{8-11}
    & $t_\text{initial}$ & $t_\text{final}$ & & $d\text{[M/H]}/dR$ & $d\text{[M/H]}/dz$ & &  $d\text{[M/H]}/dR$ & \multicolumn{3}{c}{$d\text{[M/H]}/dz$} \\
    \cline{9-11}
    & Gyr & Gyr & & dex kpc$^{-1}$ & dex kpc$^{-1}$ & & dex kpc$^{-1}$ & dex kpc$^{-1}$ & dex kpc$^{-1}$ & dex kpc$^{-1}$\\
    & & & & & & & $6<R<12$ & $2<R<6$ & $6<R<10$ & $10<R<14$ \\
 \hline
   A  & 0.0 & 8.0 & & $-0.08$ & 0.0 & & $-0.044\pm0.035$ & $+0.048\pm0.101$ & $+0.069\pm0.115$ & $+0.101\pm0.135$ \\
   A5 & 5.0 & 8.0 & & $-0.08$ & 0.0 & & $-0.059\pm0.029$ & $+0.025\pm0.090$ & $+0.030\pm0.110$ & $+0.054\pm0.126$ \\
   B  & 0.0 & 8.0 & & $-0.08$ & 0.0 & & $-0.027\pm0.031$ & $-0.198\pm0.172$ & $-0.131\pm0.105$ & $-0.118\pm0.102$ \\
  \hline
 \end{tabular}
\end{table*}

\begin{figure}
	\includegraphics[width=\hsize]{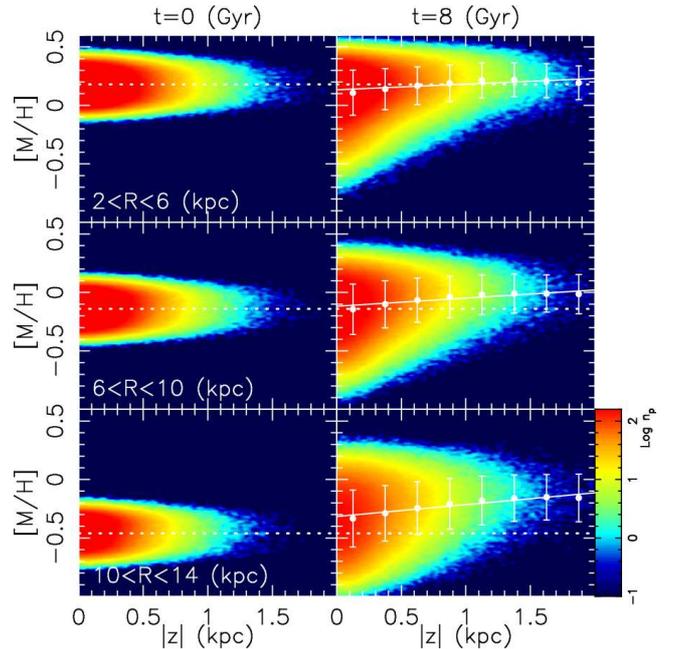}
    \caption{Vertical metallicity gradients of the thin disc component for Model~A in the radial ranges of $2<R<6$~kpc (top panels), $6<R<10$~kpc (middle panels) and $10<R<14$~kpc (bottom panels) at $t=0$ (left) and $8$ (right) Gyr. The white dots and the error bars in the right panel show the mean and dispersion of [M/H] at different height, $|z|$, and the solid line shows the best fit straight line. The dotted line shows the vertical metallicity gradient assumed at $R=4$ (top), 8 (middle) and 12 (bottom) kpc  at $t=0$ Gyr. Colour contour map indicates the number density of particles in arbitrary logarithmic units.}
    \label{fig:smapzmet-m68}
\end{figure}

Fig.~\ref{fig:smapzmet-m68} shows the vertical metallicity distribution for the thin disc particles of Model~A at $t=0$ and 8 Gyr. The evolution of the vertical metallicity gradient is summarised in Table~\ref{tab:zgrad}. Interestingly, after 8 Gyr of evolution, the vertical metallicity gradient becomes positive. This is the first key result of this paper.

\begin{figure}
	\includegraphics[width=\hsize]{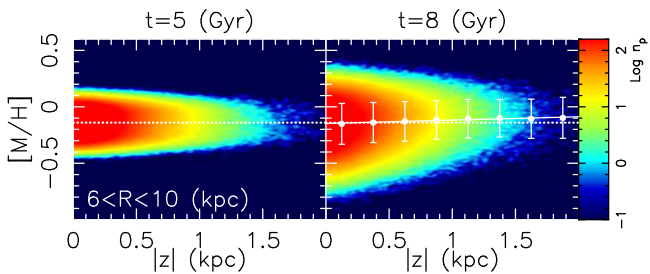}
    \caption{Same as Fig.~\ref{fig:smapzmet-m68}, but for Model A5 where the metallicity gradient was set at $t=5$ Gyr. Only the results in a radial range of $6<R<10$ kpc are shown. The dotted line shows the vertical metallicity gradient assumed at $R=8$ kpc  at $t=0$ Gyr.}
    \label{fig:smapzmet-m68t5}
\end{figure}

Fig.~\ref{fig:smapzmet-m68t5} shows the vertical metallicity distribution for the thin disc particles of Model~A5 at $t=5$ and 8 Gyr. In this Model~A5, we use the same simulation result as Model~A, but we set the metallicity gradient of $\text{[M/H]}=0.5-0.08\times~R$ with a dispersion of 0.05~dex at $t=5$~Gyr after the bar has fully developed (See Fig.~\ref{fig:snap}). This mimics 3~Gyr evolution of a mono-age thin disc population formed at $t=5$~Gyr. In this case, there is negligible heating (Fig.~\ref{fig:zprof-m68}), and the co-rotation radial migration, i.e. changes of the guiding radii with no heating, due to the bar and spiral arms is the main driver of the change of the metallicity distribution. Still, the vertical metallicity gradients become more positive, even in this short timescale. Table~\ref{tab:zgrad} presents the best fit gradient. The uncertainty of the gradient is larger than the positive gradient. However, as seen in Fig.~\ref{fig:smapzmet-m68t5}, the positive gradient is present, which is inconsistent with the observed significant negative vertical metallicity gradients as discussed in Section~\ref{sec:intro}.  

\begin{figure}
	\includegraphics[width=\hsize]{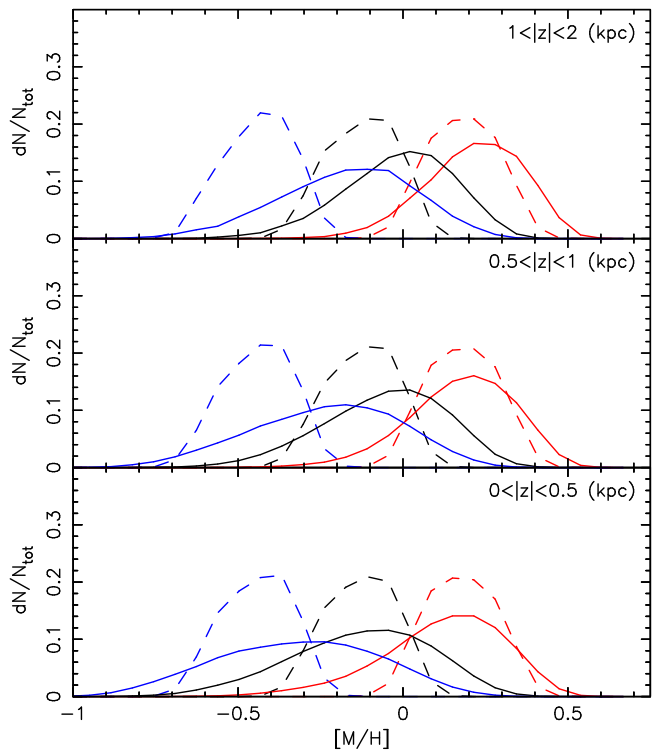}
    \caption{Metallicity distribution function (MDF) for Model~A in the radial region of $2<R<6$ kpc (red), $6<R<10$ kpc (black) and $10<R<14$ kpc (blue) at different vertical ranges of $1<|z|<2$ kpc (top), $0.5<|z|<1$ kpc (middle) and $0.0<|z|<0.5$ kpc (bottom). The solid (dashed) line shows the MDF at $t=8$ ($t=0$) Gyr.}
    \label{fig:mdfrz-m68}
\end{figure}

Fig.~\ref{fig:mdfrz-m68} shows the MDF for Model~A in the different radial and vertical regions. In the disc plane, i.e. $0<|z|<0.5$~kpc, the MDF in the radial range of $2<R<6$~kpc shows a negatively skewed MDF like what is observed in the APOGEE survey data \citep[Fig.~5 in][]{hbhnb15}. The negatively skewed MDF in the inner region of the disc indicates that a non-negligible amount of metal poor stars which were originally in the outer disc radially migrated to the inner region, and contributed to the tail of the low metallicity stars.

We note that the MDF for $10<R<14$~kpc and $0<|z|<0.5$~kpc does not show a clear positive skewness, which is inconsistent with a clear positively skewed MDF in the outer region of the Milky Way disc observed in \citet{hbhnb15}. This is partly due to our choice of the broad radial range of 4~kpc width compared to 2~kpc applied in \citet{hbhnb15}. However, we also think that this is because the bar formation leads to too large radial mixing\footnote{This may sound that the bar formation kinematically heated the thin disc too much. However, the velocity dispersion of the thin disc at $R=8.25$~kpc at $t=8$ Gyr in Model~A is $(\sigma_{\rm R}, \sigma_{\phi}, \sigma_{\rm z})=(38, 30, 18)$~km s$^{-1}$, which is comparable to what is observed for the similarly old thin disc population in the Milky Way \citep[e.g. Fig.\ 7 of ][]{hna07}. Therefore, the kinematic heating, i.e. radial scattering, due to the bar formation is not too significant in this model.}. In fact, we found a clear positive skewness in the region of $13<R<15$~kpc and $0<|z|<0.5$~kpc in Model~A5, which is more similar to what is found in \citet{hbhnb15}  \citep[see also][]{ldnhh16,mmpmp16}. Hence, we speculate that the skewness of the MDF may be higher for the stars formed after the bar formation than the stars formed before the bar formation. If we could analyse the MDF for different age groups in the outer disc, we may be able to identify the epoch of the bar formation. However, this is beyond the scope of this paper. 

\begin{figure}
	\includegraphics[width=\hsize]{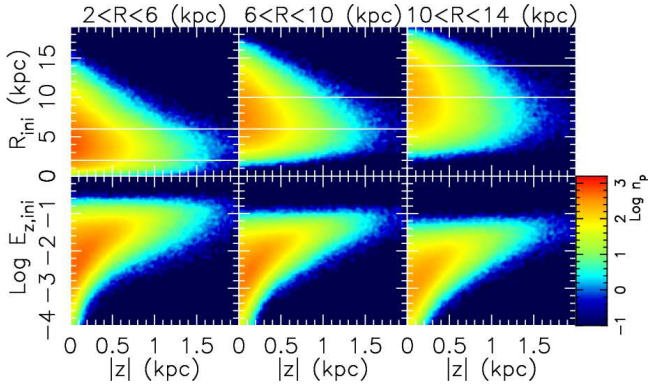}
    \caption{Top panels: Initial ($t=0$ Gyr) radius, $R_\text{ini}$, as a function of the final ($t=8$ Gyr) vertical height, $|z|$ for the particles whose final radial range is $2<R<6$ kpc (left panels), $6<R<10$ kpc (middle panels) and $10<R<14$ kpc (right panels) for Model~A. The horizontal white solid lines indicate the corresponding radial range to guide the eye. Bottom panels: Initial ($t=0$ Gyr) vertical energy, $E_\text{z,ini}$, in an arbitrary unit, as a function of the final ($t=8$ Gyr) vertical height, $|z|$, for the particles whose final radial range is $2<R<6$ kpc (left panels), $6<R<10$ kpc (middle panels) and $10<R<14$ kpc (right panels) for Model~A. Colour contour map indicates the number density of particles in arbitrary logarithmic units.}
\label{fig:reziz-m68}
\end{figure}

The MDF in the outer region, especially in the $10<R<14$~kpc samples, shows a clear trend that the peak of the MDF shifts toward higher metallicity as the height increases, which drives the positive vertical metallicity gradient seen in Fig.~\ref{fig:smapzmet-m68}. This demonstrates that the high metallicity stars which were in the inner region contribute more in the higher vertical height after they migrated outward. This is also happening even in the inner region, i.e. the stars from $R<2$~kpc contributes more at higher vertical heights. In addition, the lower metallicity stars from the outer region contributes at lower vertical heights. These combined effects drives the positive vertical metallicity gradients in the $2<R<6$~kpc sample. However, the trend is more clear in the outer region. The top panels of Fig.~\ref{fig:reziz-m68}, show the initial radius of the particles which are located in the three different radial ranges as a function of the current vertical height. In fact more particles which came from the inner region end up at the higher vertical height at every radii, which can explain the positive metallicity gradient in Fig.~\ref{fig:smapzmet-m68t5} and the MDF trend in Fig.~\ref{fig:mdfrz-m68}.

\begin{figure}
	\includegraphics[width=\hsize]{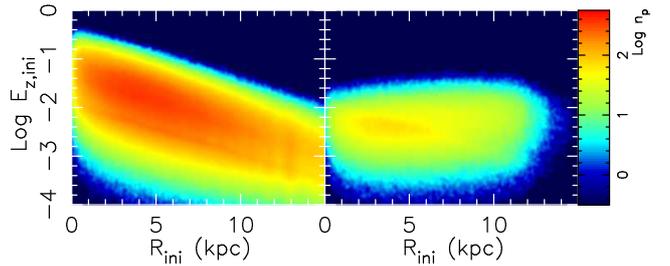}
    \caption{Initial vertical energy, $E_\text{z,ini}$, as a function of the radius, $R_\text{z,ini}$, at $t=0$ Gyr for the thin disc of Model~A (left) and the 2nd thin disc of Model~B (right). Colour contour map indicates the number density of particles in arbitrary logarithmic units.}
    \label{fig:ezri}
\end{figure}

We further investigated why more particles from the inner regions end up in the region of higher vertical height in the outer disc. We have analysed the initial vertical energy, $E_\text{z,ini}$, at $t=0$ Gyr. We first calculate the gravitational potential at the grid position in cylindrical coordinates, with logarithmic binning in the radius and linear binning in the azimuthal angle and the vertical direction from the particle distribution. The spherical NFW dark matter potential is also taken into account. Then, we measure the gravitational potential, $\Phi (R,\theta,z)$, for each particle using linear interpolation. The vertical energy, $E_\text{z,i}$ for particle $i$ is calculated following equation~(13) of \citet{sss12},
\begin{equation}
 E_\text{z,i}=\frac{1}{2} v_\text{z}^2+\Phi(R_\text{i},\theta_\text{i},z_\text{i})-\Phi(R_\text{i},\theta_\text{i},0),
\end{equation}
where $v_\text{z}$ is the vertical velocity and $\Phi(R_\text{i},\theta_\text{i},z_\text{i})$ is the gravitational potential at the particle position, $(R_\text{i},\theta_\text{i},z_\text{i})$. The bottom panels of Fig.~\ref{fig:reziz-m68} show that the particles with higher $E_\text{z,ini}$ tend to end up at higher $|z|$. The left panel of Fig.~\ref{fig:ezri} shows that at $t=0$ the thin disc particles in the inner region have higher $E_\text{z,ini}$, which is naturally expected if the vertical height of the disc is constant. This is because of the higher disc surface density and the steeper change of the gravitational potential in the vertical direction in the inner region. Therefore, if the particles which are initially in the inner region and tend to have higher $E_\text{z}$ move to the outer region, i.e. migrating outward, they reach higher vertical height and become dominant at high $|z|$ \citep{sb09b}.
Hence, if there is a negative radial metallicity gradient, the radial mixing of the particles brings more metal-rich particles from the inner disc to the higher vertical height in the outer disc, which drives a positive vertical metallicity gradient. This is inconsistent with the negative vertical metallicity gradients observed for the mono-age thin disc populations in the Milky Way as discussed in Section~\ref{sec:intro}. In the next section we propose one possibility to remedy this issue. 

Note that \citet{sss12} suggested that vertical action is conserved better than vertical energy. However, they also showed that the standard deviation of the conservation of vertical energy is about 30~\%, and it has a negligible effect on the order of magnitude difference seen in Fig.~\ref{fig:reziz-m68}. \citet{vcdna14} discussed that the particles which migrate outward were preferentially vertically colder (their mean vertical height $<|z|>$ and vertical velocity dispersion, $\sigma_\text{z}$, is lower) population, which they call "provenance bias". Therefore, after they migrate outward, they are vertically colder than the population in the outer region. This looks contradictory to our results and \citet{mfqdms12} who showed that the particles which migrate outwards are vertically hotter than the particles at their new radii. We note that Fig.~11 of \citet{vcdna14} shows the particles selected by a specific range of the guiding radius of the particles at the later epoch. We could reproduce Fig.~11 of \citet{vcdna14} with Model~A5 and also a non-barred model which is used in \citet{gkc14}, if we applied the same sampling of particles as \citet{vcdna14}. However, if we sample the particles not with the guiding radius, but with the actual radius at the final timestep, we have confirmed that the stars from the inner region are vertically hotter (their mean vertical height $<|z|>$ is higher and $\sigma_\text{z}$ is higher) than the particles which stayed within the selected radial range. Therefore, our results are consistent with these previous studies.

\subsection{Flaring Disc Model}
\label{sec:resfldm}

\begin{figure}
	\includegraphics[width=\hsize]{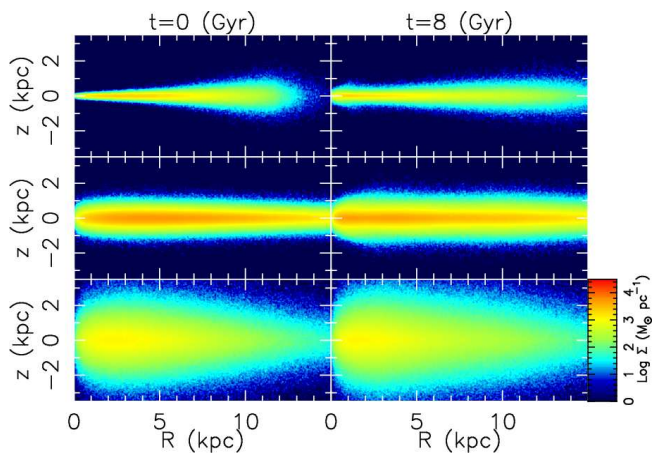}
    \caption{Particle density distribution in the $z$ vs.\ $R$ plane for the 2nd thin disc (top panels), thin disc (middle panels) and thick disc (bottom panels) components for Model~B at $t=0$ (left) and 8 (right) Gyr.}
    \label{fig:sdenRz-m94}
\end{figure}

Model~B includes a 2nd thin disc (thin2 component in Table~\ref{tab:model}) whose vertical scale height increases with radius, i.e. flaring. The 2nd thin disc component is representative of a mono-age population that is born in a flaring star-forming region, allowing us to contrast how the evolution of such a flared distribution differs from the standard thin disc stellar distribution that comprise Model~A. We embed the flared 2nd thin disc in the established thin disc. The established thin disc can be interpreted as the generations of the flared thin discs which can build up a constant scale height thin disc, if the older discs were radially smaller and thicker as will be discussed in Section \ref{sec:sum}. Ideally, the disc growth should be taken into account. However, it is numerically challenging to control the stability of the growing disc \citep[a recent heroic attempt can be found in][]{abs16}. Therefore, we ignore the growth of the disc, and add the 2nd thin disc in the established thin disc, which simplifies the numerical experiments and allows us to focus on the effects of the non-axisymmetric structures.

We tagged the metallicity at t = 0 Gyr for the 2nd thin disc particles for Model B to follow the same radial metallicity gradient as the thin disc particles for Model~A, i.e. [M/H]$=0.5-0.08\times R$ with a dispersion of 0.05 dex. We do not tag or follow the metallicity for the other components, i.e. the thin and thick disc particles, which is irrelevant for this study. The morphological evolution of the thin and thick discs were very similar to Model~A whose evolution is seen in Fig.~\ref{fig:snap}. Fig.~\ref{fig:sdenRz-m94} shows the projected density distribution for all three disc components of Model~B in the $R$-$z$ plane at $t=0$ and 8 Gyr. Our numerical experiment reveals how the vertical metallicity gradient of this flaring disc component develops after 8 Gyr of evolution, when the negative radial metallicity gradient is set at $t=0$~Gyr as described in Section~\ref{sec:meth}.

\begin{figure}
	\includegraphics[width=\hsize]{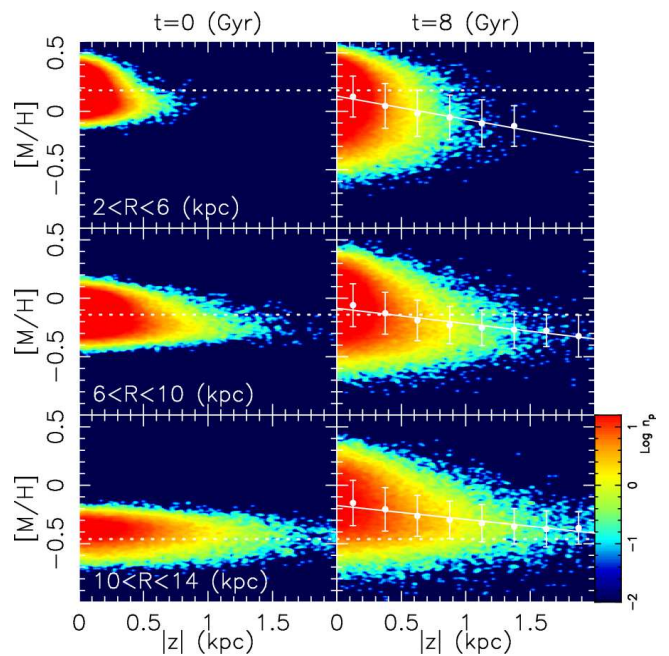}
    \caption{Same as Fig.~\ref{fig:smapzmet-m68}, but for the 2nd thin disc of Model~B.}
    \label{fig:smapzmet-m94}
\end{figure}

Fig.~\ref{fig:smapzmet-m94} shows that this model can lead to a clear negative vertical metallicity gradient for the 2nd thin disc. Therefore, if the structure of the star-forming region is a flaring disc similar to the 2nd thin disc component of Model~B, and has a negative radial metallicity gradient, the radial mixing can drive a negative vertical metallicity gradient for the stars with the same age. This is one solution to explain the observed negative vertical metallicity gradient of the sample of relatively young stars, like the ones shown in Section~\ref{sec:intro}, and solve the problem of radial mixing, which otherwise drives a positive vertical metallicity gradient. 

\begin{figure}
	\includegraphics[width=\hsize]{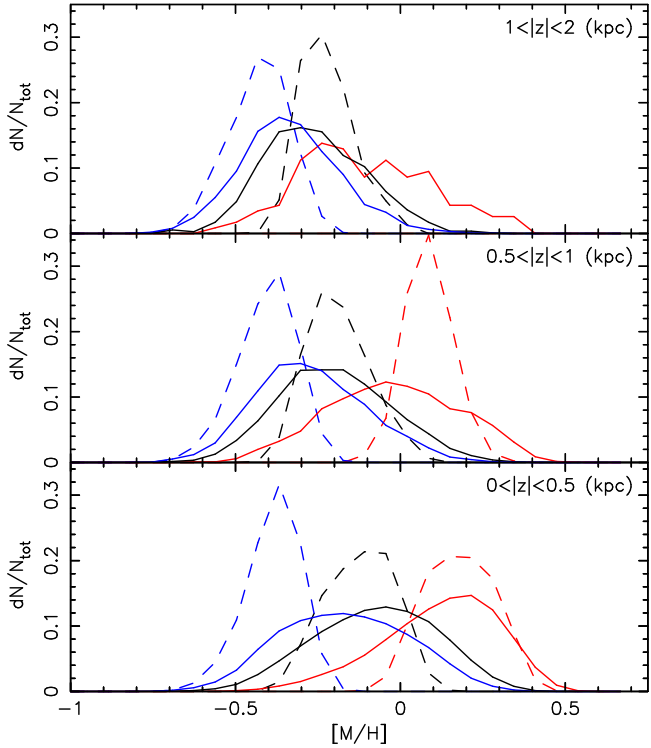}
    \caption{Same as Fig.~\ref{fig:mdfrz-m68}, but for the 2nd thin disc of Model~B. There are no particles in the region of $2<R<6$~kpc and $1<|z|<2$~kpc in the initial disc, and hence no initial MDF for this region is plotted in the top panel.}
    \label{fig:mdfrz-m94}
\end{figure} 

Fig.~\ref{fig:mdfrz-m94} shows that the peak of the MDF shifts toward lower metallicity at higher $|z|$, which drives the negative vertical metallicity gradients. Figs.~\ref{fig:smapzmet-m94} and \ref{fig:mdfrz-m94} show an interesting consequence of the flaring disc model. Because the lower metallicity population at high $|z|$ at all radii are caused by the particles from the outer region, the peak metallicities at high vertical height in $2<R<6$ and $6<R<10$ kpc are similar to the one in the outer most radial range, $10<R<14$ kpc. This leads to a steeper vertical metallicity gradient in the inner region, compared to the outer region (see also Table~\ref{tab:zgrad}), for a mono-age population formed in a flaring star forming region. This is a natural consequence of radial mixing and a flaring star-forming region, and can be tested once we have a large number of thin disc stars with accurate age measurements covering a wide radial and vertical range. 


\begin{figure}
	\includegraphics[width=\hsize]{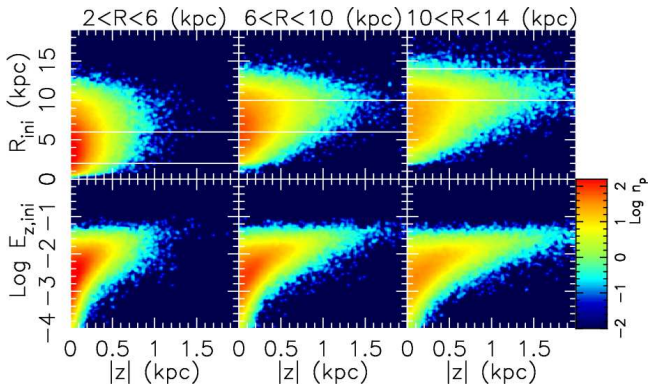}
    \caption{Same as Fig.~\ref{fig:reziz-m68}, but for the 2nd thin disc of Model~B.}
    \label{fig:reziz-m94}
\end{figure}

The upper panels of Fig.~\ref{fig:reziz-m94} show that the stars at a higher vertical height at $t=8$ Gyr originated more in the outer region, which is a clear contrast with the results in the upper panels of Fig.~\ref{fig:reziz-m68}. This drives lower metallicity at higher vertical height. The right panel of Fig.~\ref{fig:ezri} shows that there are higher $E_\text{z,ini}$ particles at the outer radii in the thin2 disc component of Model~B. As shown in the bottom panels of Fig.~\ref{fig:reziz-m94}, the higher $E_\text{z,ini}$ particles tend to end up at higher $|z|$, which explains the reason why the particles initially at the outer radii tend to have higher final $|z|$.

\section{Summary and Discussion}
\label{sec:sum}

Using several idealised N-body models of the disc galaxies similar in size to the Milky Way, our numerical experiments demonstrate that the radial mixing of stars driven by the bar and spiral arms leads to a positive vertical metallicity gradient for the same age population of the thin disc stars, if they form with a negative radial metallicity gradient. This is because when the stars in the inner region with higher metallicity migrate outwards, they tend to end up at a higher height in the outer region of the disc, due to their higher initial vertical energy. The positive vertical metallicity gradient is inconsistent with the current observational evidence in the thin disc of the Milky Way, including the negative vertical metallicity gradients that we obtained for different age ranges of the young stars in \citet{csassh16} (Sec.~\ref{sec:intro}).

We also study the evolution of the vertical metallicity gradient when the stars formed in a flaring thin disc. In this case, the stars formed in the outer disc become dominant at high vertical height at every radii, which can drive a negative vertical metallicity gradient, if there is a negative radial metallicity gradient and the stars formed in the outer region have lower metallicity.
Therefore, if a single population of the Milky Way thin disc stars is formed within a flaring star forming disc, the mono-age thin disc population can have a negative vertical metallicity gradient after a few Gyr. Our numerical experiments further predict that if a mono-age thin disc population formed in a flaring disc, the vertical metallicity gradient of the mono-age thin disc population would be steeper in the inner disc, and the peak of the MDF becomes similar at a higher vertical height, i.e. the radial metallicity gradient of the mono-age population would be shallower at a higher height. We also predict that the MDF of mono-age thin disc populations at a high vertical height would be more positively skewed in the inner disc compared to the outer disc. Interestingly, some of these trends are qualitatively consistent with the recent results from the LAMOST Spectroscopic Survey of the Galactic Anti-Centre (LSS-GAC). \citet{xlyhw15} found that all the mono-age samples (they used 2~Gyr age bin for stars younger than 8 Gyr and $8-11$~Gyr age bin) of the turn-off stars with age younger than 11 Gyr have the vertical metallicity gradients steeper in the inner radii, and the radial metallicity gradient shallower at higher vertical heights (their radial and vertical coverages are $7.5<R<13.5$~ kpc, and $|z|<2.5$~kpc, respectively).

Flaring star-forming regions are observed in the distribution of the HI and molecular gas in the Milky Way \citep[e.g.][]{stcs93,hnys16}, and it is not impossible to imagine that all the thin disc stars formed in a flaring star-forming region. In this scenario, every generation of the thin disc population are flaring and therefore the whole thin disc population is flaring, although the radial mixing will reduce the strength of the flaring, and it is likely that earlier generations of the thin disc were smaller and thicker as seen in numerical simulations \citep{bkgf04b,bkmg06,bkwgcmm13,mmssdjs15,gsgmp16}. The 2nd thin disc component of Model~B was representative of one of the many generations of the flaring discs.

The flaring younger thin disc population helps to explain the observed abundance distribution of thin and thick disc populations. Analysing a cosmological simulation result, \citet{rck14} found that the old compact thick disc population and the flaring younger thinner disc population leads to a negative radial metallicity gradient at the disc plane, and a positive radial metallicity gradient at the high vertical height, $2<|z|<3$~kpc, which is consistent with the observed radial metallicity gradients at different vertical heights in the Milky Way \citep[e.g.][]{ccz12}. \citet{rck14} predicted that if this is true, we should see a negative radial [$\alpha$/Fe] gradient at a high vertical height, because the old disc population has systematically higher [$\alpha$/Fe] than the young disc population \citep[see also][]{mpgbsb16}. This prediction has been confirmed by several observational studies. For example, from the APOGEE data \citet{acsrg14} showed almost flat radial [$\alpha$/M] gradient at $|z_\text{max}|<0.4 $~kpc, but a clear negative [$\alpha$/M] gradient at $1.5<|z_\text{max}|<3$ kpc, where $z_\text{max}$ is the maximum vertical amplitude from their orbital analysis. \citet{rck14} also predicted that there should be a negative radial age gradient for the disc stars at a high vertical hight, which should be able to be tested in the Milky Way as well as external edge-on galaxies. Furthermore, \citet{mmssdjs15} analyzed the data from two different cosmological numerical simulations, and studied the structure of the disc stars within many different age bins. They demonstrated that there is a continuous trend that younger discs are larger and thinner, but the scale-height increases with radius, i.e. flaring.
  
The recent observations also show tentative evidence of the flaring thin disc populations. From the APOGEE data, \citet{nbbah14} and \citet{hbhnb15} showed that at high-vertical height ($1<|z|<2$~kpc) in the inner region ($R<7$ kpc) the high-$\alpha$ thick disc population is dominant, and there are very few low-$\alpha$ thin disc population stars \citep[see Fig.~4 of][]{hbhnb15}. On the other hand, in the outer region ($R>9-11$ kpc), there is almost no thick disc population, while there is a dominant thin disc population at any height of the disc. This trend can be naturally explained by the compact thick disc population and a flaring thin disc population \citep[see also][]{baboym11}. Fitting the different populations of the APOGEE data with parametrised radial and vertical density profiles, \citet{brsnhsb16} also showed the compact thick disc population and flaring thin disc populations. These observational results also indicate that a flaring disc is not surprising, but rather expected. 

\citet{jibls08} found using a photometric SDSS sample that the geometrically thick disc scale length, $h=3.9$ kpc, of the Milky Way is larger than that of the geometrically thin disc, $h=2.6$ kpc. This may sound contradictory to the above discussion of a compact chemically defined thick disc population and a larger thin disc population \citep[e.g.][]{baboym11,brlhbl12,crmlb12}. However, \citet{jibls08} studied geometrically thick and thin discs which are defined purely by the spatial distribution of stars, but not by age or chemical abundances. \citet{mmssdjs15} discussed that even if the chemically decomposed thick disc population is more compact than the thin disc population, the flaring thin disc population can contribute to the geometrically thick disc at the outer radii \citep[see also][]{rck14}, and lead to a larger geometrically defined thick disc than the chemically defined thick disc population. In addition, the overall stellar distribution does not have to show a clear flaring, because the scale-height of the geometric thick disc structure is determined by the mixture of the thin and thick disc populations. The flaring thin disc population would be clearly identified, only when the populations are decomposed by the metal abundances or age. 

  The European Space Agency (ESA)'s \textit{Gaia} mission will soon provide us the 6D (or 5D for faint stars) phase space distribution of hundreds of  million of giant stars which cover a large volume of the Galactic discs \citep{rlrig12}. The \textit{Gaia} data will be supplemented with the chemical abundance information from high-resolution spectroscopic surveys, such as Gaia-ESO, APOGEE and GALAH, which are also possible to provide the age of stars \citep[e.g.][]{nhrmph16}. In addition, the K2 mission with \textit{Kepler} will provide the accurate ages of giant stars in several different directions of the Galactic discs \citep[e.g.][]{shsjl15}. Ultimately, ESA's \textit{PLATO} mission will uncover the age of bright stars in a large fraction of the sky. These data will tell us the structure and metallicity distribution of the mono-age Galactic disc populations, which will provide us with strong constraints on the formation scenario and subsequent evolutionary history of the Galactic disc.

 \section*{Acknowledgments}
We thank an anonymous referee for their constructive comments and helpful suggestions which have improved the manuscript. DK also thank Ralph Sch\"onrich for fruitful discussion.
DK and JASH gratefully acknowledge the support of the UK's Science \& Technology Facilities Council (STFC Grant ST/K000977/1 and ST/N000811/1).  RJJG acknowledges support by the DFG Research Centre SFB-881 'The Milky Way System' through project A1. The calculations for this paper were performed on the UCL facility Grace, the IRIDIS HPC facility provided by the Centre for Innovation and the DiRAC Facilities (through the COSMOS and MSSL-Astro consortium) jointly funded by STFC and the Large Facilities Capital Fund of BIS.  We also acknowledge PRACE for awarding us access to their Tier-1 facilities.




\bibliographystyle{mnras}







\bsp	
\label{lastpage}
\end{document}